\documentstyle[epsf,aps,twocolumn]{revtex}            

\def\lesssim{\mathrel{\hbox{\rlap{\hbox
{\lower4pt\hbox{$\sim$}}}\hbox{$<$}}}}
 
\def\gtrsim{\mathrel{\hbox{\rlap{\hbox
{\lower4pt\hbox{$\sim$}}}\hbox{$>$}}}}

\begin{document}
\draft
\title{The Formation of Supermassive Black Holes and the Evolution
of Supermassive Stars}
\author{Kimberly C. B. New}
\address{Los Alamos National Laboratory, Los Alamos, NM 87545}
\author{Stuart L. Shapiro}
\address{Department of Physics, Department of Astronomy, and
National Center for Supercomputing Applications, University
of Illinois at Urbana-Champaign, Urbana, IL 61801}

\date{\today, current draft}

\maketitle
    \begin{abstract}
The existence of supermassive black holes is supported by a
growing body of observations. Supermassive black holes
and their formation
events are likely candidates for detection by proposed long-wavelength,
space-based gravitational wave interferometers like LISA.  However, the nature
of the progenitors of supermassive black holes is rather uncertain.
Supermassive black hole formation scenarios that
involve either the stellar dynamical evolution
of dense clusters or the hydrodynamical evolution of
supermassive stars have been proposed.  Each of these formation
scenarios is reviewed and the evolution of supermassive stars
is then examined in some detail.  Supermassive stars
that rotate uniformly during their secular cooling phase will
spin up to the mass-shedding limit and eventually contract to the point of
relativistic collapse.  Supermassive stars
that rotate differentially as they cool will likely
encounter the dynamical bar mode instability prior to the onset of relativistic
collapse.  A supermassive star
that undergoes this bar distortion, prior to or during collapse,
may be a strong source of
quasiperiodic, long-wavelength gravitational radiation.
    \end{abstract} 

\section{Supermassive Black Holes}

There is a large body of observational evidence that supermassive black holes
(SMBHs) exist in the centers of many, if not most galaxies (see e.g., the
reviews of Rees \cite{rees98} and Macchetto \cite{macc99}).
The masses of SMBHs in the centers of more than 45 galaxies
have been estimated from observations \cite{feme00} and there are more
than 30 galaxies in which the presence of a SMBH has been confirmed
\cite{korm00}.  The properties of a few confirmed SMBHs are given in
Table 1.

Several properties have been deduced from these observations
\cite{korm00,feme00,gebh00,haka00,wang00,mefe00}.  A correlation between
the SMBH mass and both the mass and the velocity dispersion
of the bulge of the host galaxy is observed.  These results
suggest that the formation and evolution of SMBHs and the bulge
component of their
host galaxies may be closely related.  The largest SMBHs are found
in elliptical galaxies and thus may result from galaxy mergers,
as ellipticals themselves are thought to form via merger.

Because of their size and mass, SMBHs are expected to be sources
of long-wavelength, low-frequency gravitational radiation.  Thus,
phenomena involving SMBHs might be detectable with proposed
space-based gravitational wave detectors, like the Laser Interferometer
Space Antenna (LISA) \cite{folk98}.  For example, LISA might be
able to detect the collapse of a supermassive star to a
SMBH, the coalescence of two SMBHs, or the coalescence of a compact
star and a SMBH \cite{thbr76,thor95,hibe95,sire97,schu97}.
The event rates of these phenomena are uncertain due to the uncertainty
in the nature of the progenitors of SMBHs.  However, the rates
could be appreciable because SMBHs are present in
many galaxies.

\subsection{Formation Mechanisms}

The possible formation mechanisms for SMBHs involve either the
stellar dynamics of dense star clusters or the hydrodynamics of
supermassive stars (SMSs).

\subsubsection{Stellar Dynamical Routes}

In general terms, dense star clusters are formed via a conglomeration
of stars, produced by fragmentation of the primordial gas.  In one
cluster scenario for SMBH formation, massive stars form via stellar collisions
and mergers in the cluster and then evolve into stellar-mass black
holes.  The merger of these holes, as they grow and settle to the
cluster center, leads to the build up of one or more SMBHs
\cite{qush90,lee95,hibe95}.

An alternative cluster scenario centers on
the catastrophic collapse of the cluster core.  Shapiro and Teukolsky
have numerically demonstrated that cluster cores can become dynamically
unstable to relativistic collapse \cite{shte85a}.  They also showed
that if the core is made up of compact stars, it will evolve to the
onset of collapse on a relaxation timescale \cite{shte85b}.  As the
compact core evolves to this point of relativistic instability, the mass
of its black holes will increase through collisions and gravitational
radiation driven mergers.  This will accelerate the approach to relativistic
instability, leading to SMBH formation \cite{zepo65,qush87}.

\subsubsection{Hydrodynamical Routes}

SMSs may contract directly out of the primordial gas, if radiation
and/or magnetic field pressure prevent fragmentation
\cite{hare93,eilo95,haeh98}
(see also \cite{lora94,brom99,abel00}).  Alternatively,
they may build up from fragments of stellar collisions in clusters
\cite{sand70,bere78}.

The evolution of SMSs will ultimately lead to the onset of relativistic
instability \cite{iben63,chan64a,chan64b,shte83}.  If
the mass of the star exceeds $10^6 M_{\odot}$, the star will then
collapse and possibly form a SMBH.  If the star is less massive,
nuclear reactions may lead to explosion instead of collapse.

\section{Supermassive Stars}

Identifying the scenario by which SMBHs form is of fundamental
importance to a number of areas of astrophysics.
The remainder of this paper examines one of the
possible progenitors of SMBHs discussed above, SMSs.

Supermassive stars are radiation dominated, isentropic, and
convective \cite{zeno71,shte83,lora94} and are thus well
represented by a $n=3$ polytrope.  As mentioned previously,
if the star's mass exceeds $10^6 M_{\odot}$, nuclear burning
and electron/positron annihilation are not important.

After formation, a SMS will evolve through a phase of
quasistationary cooling and contraction.  If the SMS is
rotating when it forms, conservation of angular momentum
requires that it spin up as it contracts.  The evolutionary
path taken by the SMS during this cooling phase depends on
the strength of its viscosity and magnetic fields and on
the nature of its angular momentum distribution.

\subsection{Cooling Evolution of Uniformly Rotating Supermassive Stars}

There are two possible evolutionary regimes for a cooling SMS.
In the first regime, viscosity or magnetic fields are strong
enough to enforce {\it uniform} rotation throughout the star as it
contracts.

Baumgarte and Shapiro \cite{bash99} have recently studied the
evolution of a uniformly rotating SMS up to the onset of
relativistic instability.
They demonstrated that a uniformly rotating,
cooling SMS will eventually spin up to its mass shedding limit.
The mass shedding limit is encountered when matter at the
star's equator rotates with the Keplerian velocity.  The limit
can be represented as a ratio $\beta_{shed}=(T/|W|)_{shed}$
of the star's rotational kinetic energy $T$ to gravitational potential
energy $W$.  In this case, $\beta_{shed}=9\times 10^{-3}$.
The star will then evolve along a mass shedding sequence, losing
both mass and angular momentum.  It will eventually contract
to the point of relativistic instability.

Baumgarte and Shapiro used both a 2nd order,
post-Newtonian approximation and a fully general relativistic
numerical code to determine that the onset of the instability
occurs at a ratio of $R/M \sim 450$, where $R$ is the \\ star's radius.
Here and throughout this paper $G=c=1$.
Note that a 2nd order, post-Newtonian approximation was needed
because rotation stabilizes the destabilizing role of nonlinear
gravity at the 1st post-Newtonian level.

The major result of Baumgarte and Shapiro's work is that the following
universal ratios exist for the critical configuration at the onset of
relativistic instability:  $T/|W|$, $R/M$, and $J/M^2$. Here $J$ is
the total angular momentum of the star.  These ratios are completely
independent of the mass of the star or its prior evolution.  Because
uniformly rotating SMSs will begin to collapse from a universal
configuration, the subsequent collapse and the resulting gravitational
waveform will be unique.

\subsubsection{Collapse Outcome and Gravitational Radiation Emission}
The outcome of SMS collapse can only be determined with numerical,
fully relativistic three-dimensional hydrodynamics simulations.
To date, such simulations have only been published for nearly
spherical collapse.  The numerical simulations of Shapiro and
Teukolsky \cite{shte79} demonstrate that this type of collapse
is nearly homologous.

In this case the collapse time $\tau_{coll}$
is roughly the free-fall time at the horizon (where $R=2M$)
\begin{eqnarray}
\tau_{coll} &=& \biggl(\frac{R^3}{4\pi M}\biggr)^{1/2} \nonumber \\
            &=& 14\, {\rm sec}\, [M/10^6 M_{\odot}]^{-1}.
\end{eqnarray}
The peak gravitational wave frequency $f_{GW}=\tau_{coll}^{-1}$
is then $10^{-2}\, {\rm Hz}$, if the mass of the star is $10^6 M_{\odot}$.
This is in the middle of LISA's frequency band of $10^{-4}-1\, {\rm Hz}$
\cite{thor95,folk98}.

The amplitude $h$ of this burst signal can be estimated roughly in
terms of the star's quadrupole moment
\begin{eqnarray}
h &\leq& \epsilon \frac{2M^2}{Rd} \nonumber \\
  &\leq& \epsilon \, 5\times 10^{-17}\, [M/10^6 M_{\odot}][d/1{\rm Gpc}]^{-1}.
\end{eqnarray}
Here $d$ is the distance to the star and $\epsilon \sim T/|W|$ is a measure of
the star's deviation from spherical symmetry.  In this
case, $\epsilon$ will be much less than one even near the horizon, since
the collapse is nearly spherical.

Three-dimensional, general relativistic simulations of nonspherical,
rotating collapse are underway \cite{shib00}.
The results are not available as yet.  However, there are two 
possible outcomes of this type of collapse that can be discussed.

The first outcome is direct collapse to a SMBH, from the onset of instability.
In this case $\epsilon$
will be on the order of one near the horizon.  Thus the peak amplitude
(see equation 2) of the burst signal will be $h \sim 5\times 10^{-17}\,
[M/10^6 M_{\odot}][d/1{\rm Gpc}]^{-1}$.  

Alternatively, the star may encounter the dynamical bar mode instability.
The bar mode is the strongest of a set of global nonaxisymmetric
instabilities that may be encountered by a rapidly rotating object.
This instability will deform the star into a bar-shaped configuration,
making it a strong source of long-wavelength, quasiperiodic gravitational
radiation.  During the development of the instability, angular momentum
and mass will be transported outwards \cite{tohl85,new00,brow00}.
This could hasten the star's eventual collapse.  Previous linear
and nonlinear analyses indicate that $\beta_{bar} \sim 0.27$ is likely
to be an upper limit for the onset of this instability, for a wide range
of polytropic equations of state and rotation laws (see, e.g.,
\cite{pick96,toma98}).

Baumgarte and Shapiro \cite{bash99} have estimated that a uniformly
rotating SMS will reach $\beta$$\sim$$0.27$ when $R/M$=15.  This
estimate awaits confirmation from three-dimensional, general
relativistic hydrodynamical simulations.

The frequency of the quasiperiodic gravitational radiation emitted
by the bar can be estimated in terms of its rotation frequency to be
\begin{eqnarray}
f_{GW} &=& 2 f_{bar} \sim 2\biggl(\frac{GM}{R^3}\biggr)^{1/2}
\nonumber \\
       &\sim& 2 \times 10^{-3}\, {\rm Hz} \, [M/10^6M_{\odot}]^{-1},
\end{eqnarray}
when $R/M$=15.   The corresponding amplitude of the radiation,
again estimated in terms of the star's quadrupole moment, is
\begin{eqnarray}
h &\sim& \frac{2M^2}{Rd} \nonumber \\
  &\sim& 6\times 10^{-18}\, [M/10^6 M_{\odot}][d/1{\rm Gpc}]^{-1}.
\end{eqnarray}
The emission of gravitational radiation will continue as long as
the collapsing SMS's bar shape persists.

\subsection{Cooling Evolution of Differentially Rotating Supermassive Stars}

In the opposite evolutionary regime, neither
viscosity nor magnetic fields are strong
enough to enforce uni\-form rotation throughout the cooling SMS as it
contracts.  In this case it has been shown that the angular momentum
distribution is conserved on cylinders during contraction
\cite{boos73}.  Because viscosity and magnetic fields are weak,
there is no means of redistributing angular
momentum in the star.  So even if the star starts out rotating
uniformly, it cannot remain so.

The star will then rotate {\it differentially} as it cools and contracts.
In this case, the subsequent evolution de\-pends on the star's initial
angular momentum distribution, which is largely unknown.

One possible outcome is that the star will spin up to mass shedding
(at a different value of $\beta_{shed}$ than a \\ uniformly rotating
star) and then follow an evolutionary path that may be similar to
that described by Baumgarte and Shapiro \cite{bash99}.

The alternative outcome is that the star will encounter the dynamical
bar instability prior to reaching the mass shedding limit.  As discussed
above, the bar shape induced by this instability will make the star
a source of quasiperiodic, long-wavelength gravitational radiation.
The outward transport of angular momentum that occurs during the instability
could hasten the eventual collapse of the star.

New and Shapiro \cite{nesh00} have investigated the evolution of
differentially rotating SMSs.  Because the angular momentum distribution
of these stars is unknown, they examined SMS models with several different
initial angular momentum distributions.  The goal of their work was
to determine if the bar instability or mass shedding limits are reached
by these SMS models.

New and Shapiro's strategy was to examine equilibrium sequences of
SMS models, each of which was constructed with a different rotation
law.  The individual models on each sequence were constrained to have
the same $M$ and $J$, since these quantities are conserved during the
cooling evolution of a SMS.  However, the models along a sequence
have decreasing entropy and thus decreasing axis ratio $R_p/R_e$,
where $R_p$ is the polar radius and $R_e$ is the equatorial radius.
A sequence is thus representative of the quasistatic, cooling/contracting
evolution of a {\it single} SMS.  This quasistatic approximation
is appropriate because the cooling timescale is much longer than
the hydrodynamic timescale for $M \lesssim 10^{13} M_{\odot}$ \cite{bash99}.
New and Shapiro examined each of the sequences to determine
whether the limits $\beta_{bar}$ or $\beta_{shed}$ 
are reached for a SMS model with the given rotation law. 

All of the sequences New and Shapiro examined were constructed
with Hachisu's self-consistent field (HSCF) technique \cite{hach86}.
The HSCF method builds individual models in hydrostatic equilibrium,
such that their pressure, gravitational, and centrifugal forces are
in balance.

The HSCF method requires the choice of a barotropic equation of
state $P=P(\rho)$.  All of the SMS models examined by New and Shapiro
were constructed with a polytropic equation of state for which
\begin{equation}
P=K\rho^{1+1/n}.
\end{equation}
As mentioned above, the structure of a SMS is well represented by
an $n=3$ polytrope.  In this equation of state, the polytropic constant
$K$ is a measure of the specific entropy of the model.

The selection of appropriate rotation laws was constrained by the
fact that each rotation law chosen had to enforce conservation
of the contracting star's specific angular momentum profile
\cite{boos73}.

One rotation law that satisfies this constraint
is the so-called $n'$=3 law \cite{boos73}.  Its specific angular
momentum profile $j(m)$ is 
\begin{equation}
j(m)=a_1+a_2\biggl(1-\frac{m(\varpi)}{M}\biggr)^{\alpha_2}
          +a_3\biggl(1-\frac{m(\varpi)}{M}\biggr)^{\alpha_3},
\end{equation}
where $M$ is the total mass of the system, $m$ is the mass interior
to cylindrical radius $\varpi$, and the numerically
determined constants are $a_1$=13.27, $a_2$=163.3, $a_3$=-176.5,
$\alpha_2$=0.2353, and $\alpha_3$=0.2222.
This angular momentum profile is identical to that of a uniformly
rotating, spherical $n$=3 polytrope.  Thus an equilibrium sequence
constructed with this rotation law is representative of the evolution
of a SMS that rotates uniformly {\it prior} to its cooling phase.
New and Shapiro \cite{nesh00} have constructed an equilibrium sequence
with this rotation law.

Because the initial rotation profiles of SMSs are unknown,
New and Shapiro also examined sequences constructed
by Hachisu, Tohline, and Eriguchi (\cite{hach88}; hereafter, HTE).
These sequences were built with the following parameterized
angular momentum profile:
\begin{equation}
j(m)=(1+q)(J/M)\bigl[1-(1-m(\varpi)/M)^{1/q}\bigr].
\end{equation}
Here, the index $q$ specifies the rotation law.  Note that the
limiting case of $q$=0 corresponds to the j-constant
rotation law.

Nearly spherical models built with these ``q-laws'' are differentially
rotating.  Thus, HTE's $n$=3, q-law sequences are
representative of the evolution of SMSs, with a wide range of
{\it initial differential} rotation profiles. 

\subsubsection{Evolutionary Scenarios}
A detailed discussion of the properties of the n'=3 sequence,
constructed by New and Shapiro, and the q-law sequences of HTE can
be found in \cite{nesh00}. In what follows we summarize these
properties as they relate to the evolutionary scenarios of
differentially rotating SMSs.

Density contour plots of selected models from the $n'$=3 sequence
are shown in Figure 1.
The $n'$=3 sequence terminates due to mass shedding at
$\beta_{shed} \gtrsim 0.30$.  $\beta_{shed}$ exceeds the
likely upper limit for the bar instability $\beta_{bar}\lesssim 0.27$.
Thus we expect that a SMS with this rotation law should never reach
the mass shedding limit, but will instead encounter
the dynamical bar instability near $R_p/R_e \sim 0.004$. This is the axis ratio
of the model with $\beta = 0.27$ (see Figure 1d). 

No mass shedding limits exist on HTE's q-law sequences.  Each
of these sequences makes a continuous transition from spheroidal
to toroidal configurations
at values of $\beta_{trans}>0.33>0.27\gtrsim\beta_{bar}$.
Thus, $n$=3 models with these
$q$-indexed laws would likely encounter the bar mode as spheroids.

We note that the hydrodynamical study of \cite{pick96} indicates that
$\beta_{bar}$ may be less than $0.27$ for rotation laws that
place a significant amount of angular momentum in equatorial mass
elements.  Their results predict that the $m$=2 stability limit may
be less than $\sim 0.20$ for models with the $n'$=3 rotation law.
Their simulations, of $n$=1.5 polytropes, also suggest that a
one-armed spiral, $m$=1 mode may become increasingly dominant over
the $m$=2 mode as the equatorial concentration of angular momentum increases.
Note that the grid resolutions used in \cite{pick96}
were likely not sufficient to accurately model the development of
instabilities in models with these extreme differential rotation
laws \cite{toma98}.  However, the results of \cite{pick96} and
the linear and nonlinear stability analyses of \cite{toma98}
confirm that $0.27$ is likely to be an approximate upper limit
to $\beta_{bar}$, for a variety of polytropic indices and rotation
laws.  The analysis presented in \cite{nesh00}
assumes that the $m$=2 bar mode is the dominant mode and that
$\beta_{bar}\lesssim 0.27$.

Even if the actual value of $\beta_{bar}$ is less than
$0.27$, the qualitative nature of the results of New and Shapiro
would not change.
That is, the sequences they examined would still have models that
are unstable to the bar mode.  In addition, their quantitative estimates
of the characteristics of the gravitational radiation emission \\
presented below would only change by a numerical factor of order
1-10, even if $\beta_{bar}$ were as low as $0.22$.

\subsubsection{Gravitational Radiation Emission}
The results of New and Shapiro \cite{nesh00} indicate
that a bar mode phase is likely to be encountered by differentially
rotating SMSs with a wide range of initial angular momentum distributions. 
Hydrodynamical simulations are needed to follow the evolution 
of the star as the instability develops, to compute the
gravitational radiation waveforms emitted, and to determine the
fate of a SMS that undergoes the bar instability.

Previous hydrodynamical simulations of the bar instability in
models with other rotation laws and a different polytropic equation
of state indicate that the outcome of this instability is a
persistent bar-like structure that emits quasiperiodic gravitational
radiation over many cycles \cite{shib00,new00,brow00,saij00}.
If a similar outcome
re\-sults from the bar instability in a SMS, the quasiperiodic,
long-wavelength gravitational radiation emitted could be detected
by LISA.

The frequency of these quasiperiodic gravitational waves can be
estimated from the expected bar rotation rate $\Omega_{bar}$.
The model on New and Shapiro's $n'$=$3$ sequence with
$\beta$$\sim$$\beta_{bar}$=$0.27$
has a central rotation rate $\Omega_{c}$=$1.02\times 10^{-1}
{\rm Hz}\, [M_6 \beta_{-5}^{-3} R_{17}^{-3}]^{1/2}$.  Here
$M_6\equiv M/10^6 M_{\odot}$,
$R_{17}\equiv (R_e)_0/10^{17}\,{\rm{cm}}$,
and $\beta_{-5}\equiv \beta_0/10^{-5}$.
The subscript 0 denotes the value is for
the (nearly) spherical progenitor star, prior to the start of its
cooling phase.  In previous hydrodynamics 
simulations of the bar mode
instability \cite{new00}, $\Omega_{bar}$ was $\sim 0.4 \Omega_{c}$.  With
this relation between $\Omega_{bar}$ and $\Omega_{c}$, the gravitational
wave frequency $f_{GW}$ can be estimated to be
\begin{eqnarray}
f_{GW} &=& 2 f_{bar} = 2 \frac{\Omega_{bar}}{2\pi} \sim \frac{0.4}{\pi}\Omega_{c} \nonumber \\
&\sim& 1\times 10^{-2} {\rm Hz}\, [M_6 \beta_{-5}^{-3} R_{17}^{-3}]^{\frac{1}{2}
}.
\end{eqnarray}
For a SMS of $10^6 M_{\odot}$, which begins as a slowly rotating star of radius
$10^{17} {\rm cm}$ with $\beta_0$=$10^{-5}$, this yields a frequency of
$1\times 10^{-2} {\rm Hz}$. This frequency
is in the range in which LISA is expected to be most sensitive,
$10^{-4}$-$1 {\rm Hz}$.
The choice $\beta_0$=$2\times 10^{-4}/R_{17}$ is the maximum value of $\beta_0$
for which $f_{GW}$ (=$10^{-4} {\rm Hz}$) is still in LISA's
range of sensitivity.

The strength of the gravitational wave signal can be estimated roughly to be
\begin{eqnarray}
h &\sim& \frac{G}{c^4} \frac{\ddot{Q}}{d} \sim \frac{G}{c^4}
\frac{M R_{bar}^2 f_{bar}^2}{d} \nonumber \\
  &\sim& 4 \times 10^{-15} \biggl(\frac{d}{1\, {\rm Gpc}}\biggr)^{-1}
M_6^2 \beta_{-5}^{-1} R_{17}^{-1},
\end{eqnarray}
where $\ddot{Q}$ is the second time derivative of the star's
quadrupole moment and $d$ is the distance, which we \\ scale to 1 Gpc
(the Hubble distance is $\sim$ 3 Gpc).  Here we have used $R_{bar}
\sim R_e$=$4.14\times 10^{13} \beta_{-5} R_{17}\,{\rm{cm}}$ for the
$n'$=$3$ model of \cite{nesh00} that reaches the point $\beta\sim\beta_{bar}$.

The bar will decay on a secular
timescale due to dissipative effects.  For differentially rotating
SMSs, the largest source of dissipation will be gravitational radiation.
The gravitational radiation damping timescale $\tau_{GW}$
is approximately
\begin{equation}
\tau_{GW} \sim \frac{T}{\bigl(\frac{dE}{dt}\bigr)_{GW}},
\end{equation}
where $T$ is the rotational kinetic energy and
$(dE/dt)_{GW}$ is the rate at which gravitational radiation carries
energy away from the system.  Recall that $T=\beta |W|$.
The gravitational potential energy $|W| \sim GM^2/R_{bar}$.  Thus,
\begin{equation}
T=\beta \frac{G M^2}{R}.
\end{equation}
The radiation rate can be estimated as  \cite{shte83}
\begin{equation}
\bigl(\frac{dE}{dt}\bigr)_{GW} \sim \frac{G}{c^5} \frac{M}{R^2} v^6.
\end{equation}
In this case the characteristic velocity of the system is
$v=(\beta G M/R)^{1/2}.$  Substitution of equations (11) and (12)
into equation (10) yields
\begin{eqnarray}
\tau_{GW} &\sim& \frac{c^5}{G^3} \frac{R_{bar}^{4}}{\beta^2 M^3} \nonumber \\
          &\sim& 1 \times 10^4 \, {\rm yrs} \, M_6^{-3} [\beta_{-5} R_{17}]^4.
\end{eqnarray}

The number of cycles $\cal{N}$ for which the signal will persist is
\begin{eqnarray}
\cal{N}&\sim&\tau_{GW} f_{GW} \nonumber \\
 &\sim& 4 \times 10^9 [M_6 \beta_{-5} R_{17}]^{-5/2}.
\end{eqnarray}
The quasiperiodicity of such a signal will assist in its detection
\cite{schu97}.

The fraction of the mass $(dM/M)_{GW}$ radiated via gravitational radiation over
the interval $\tau_{GW}$ can be estimated as 
\begin{eqnarray}
\biggl(\frac{dM}{M}\biggr)_{GW} &\sim& \tau_{GW}
 \bigl(\frac{dE}{dt}\bigr)_{GW} \frac{1}{Mc^2} \nonumber \\
        &\sim& \frac{G}{c^2} \frac{\beta M}{R} \nonumber \\
        &\sim& 1 \times 10^{-3} M_6 [\beta_{-5} R_{17}]^{-1}.
\end{eqnarray}

\section{Conclusions and Future Work}

There is strong evidence that SMBHs exist.  However, the nature
of their progenitors is uncertain.  Proposed formation mechanisms
involve the evolution of stellar clusters or SMSs.

Recent studies indicate that differentially
rotating SMSs, and possibly uniformly rotating SMSs as well,
are likely to encounter the dynamical bar instability
and thus emit quasiperiodic, long-wavelength gravitational radiation.
SMSs will also emit burst gravitational wave signals as they undergo
relativistic collapse.

Linear and nonlinear stability analyses are needed to
determine precisely the onset of the $m$=2 dynamical bar instability
for the SMS models considered here.  Such analyses are
also needed to determine the relative importance of various unstable
nonaxisymmetric modes in SMS models (as a dominant $m$=1 mode would
change the characteristics of the gravitational radiation emission).

In addition, three-dimensional hydrodynamical simulations
are necessary to study the evolution of SMSs as they undergo
the bar instability and/or relativistic collapse,
to compute the gravitational waveforms
emitted, and to determine the final fate of the star.

Future investigations involving hydrodynamical simulations of
SMS models would benefit from improved knowledge of initial conditions,
such as an appropriate value for $\beta_0$.  These appropriate
initial conditions could be determined from studies of large-scale
structure and cosmology.

Most importantly, studies are required to assess the roles of viscosity
and magnetic fields in rotating SMSs to judge whether they are sufficient
to drive these configurations to uniform rotation prior to bar instability.

\acknowledgments
This work has been supported in part by NSF Grants AST 96-18524
and PHY 99-02833 and NASA Grants NAG5-7152 and NAG5-8418 to the
University of Illinois at Urbana-Champaign.
A portion of this work performed under auspices of the U.S. Department of Energy
by Los Alamos National Laboratory under contract W-7405-ENG-36.




%
%
%
%


\begin{figure}
\caption{Snapshots of a cooling SMS.
Density contours of \\ selected models on the
$n'$=3 equilibrium sequence are shown in the
$(x>0,z>0)$ plane.  The maximum density is normalized
to unity.  The highest density contour level is 0.9; subsequent
contour levels range from $10^{-1}$ to $10^{-10}$ and are
separated by a decade.  The axis ratios $R_p/R_e$ of the models displayed
are (a) 1.00; (b) 0.700; (c) 0.300; (d) 0.004; (e) 0.002.
The model with $R_p/R_e$=0.004 shown in (d) has $\beta$=$\beta_{bar}
\sim 0.27$.}
\label{den}
\end{figure}

\begin{table}
\begin{center}
\begin{tabular}{cccc}
Galaxy &  Mass ($M_{\odot}$) & Radius (pc) & Reference \\
\tableline
NGC 4258 & $\sim 3.6 \times 10^7$ & $<13$ & \cite{miyo95} \\
M87 & $\sim 3.2 \times 10^9$ & $<3.5$ & \cite{macc97} \\
NGC 4151 & $\sim 10^9$ & $<60$ & \cite{wing99} \\
Milky Way & $2.6-3.3\times 10^6$ & $<8$ & \cite{genz00}\\
\end{tabular}
\end{center}
\caption{The host galaxies, masses, and radii of selected confirmed SMBHs.}
\label{table 1}
\end{table}

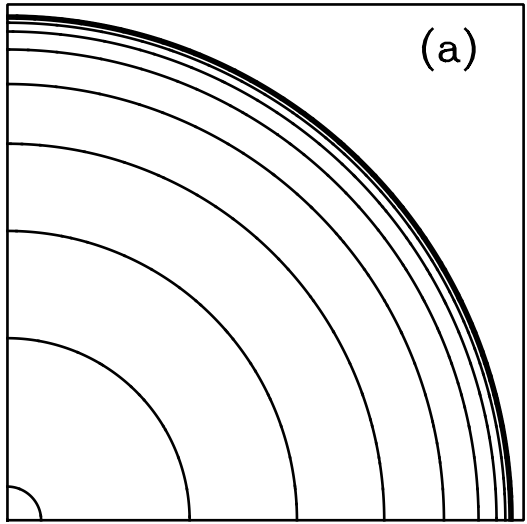
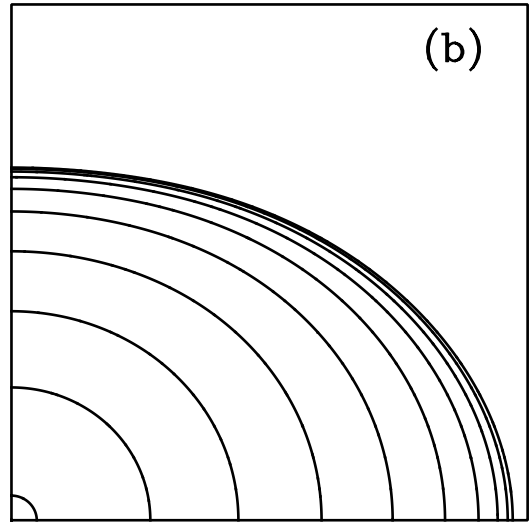
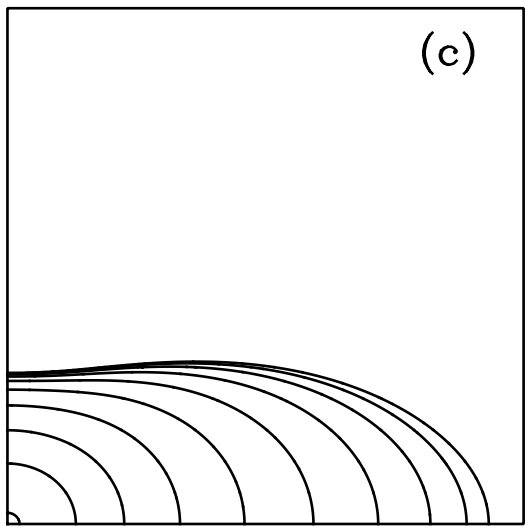
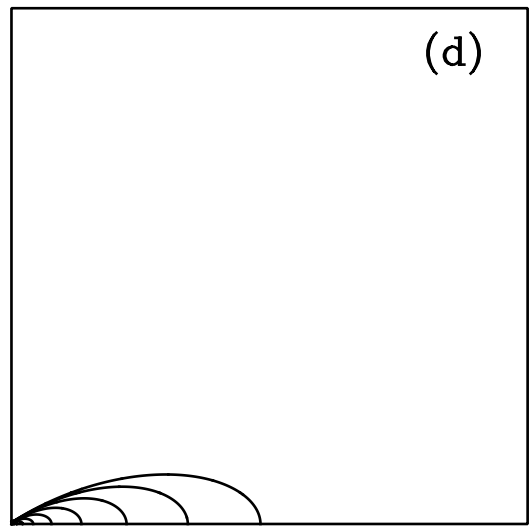
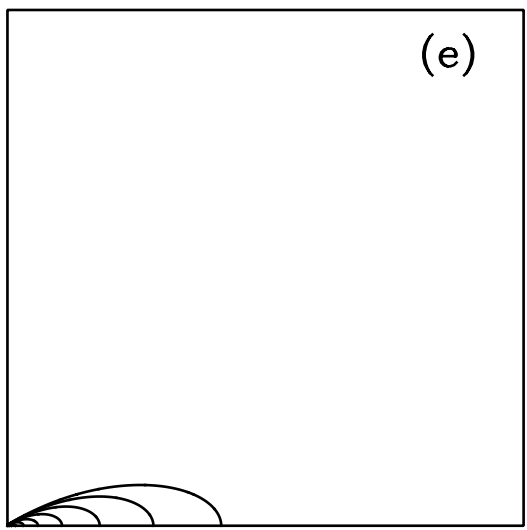
\end{document}